\documentstyle[epsfig,prb,aps]{revtex}
\tightenlines

\def\beginwide{
        \end{multicols} \vspace*{-0.5cm} \noindent
        \rule{3.5in}{.1mm}\rule{.1mm}{5mm} \widetext \medskip }
\def\beginwidetop{
        \end{multicols} \vspace*{-0.5cm} \noindent
        \widetext \medskip }
\def\endwide{
        \hspace*{3.35in}~\rule[-5mm]{.1mm}{5mm}\rule{3.5in}{.1mm}
        \begin{multicols}{2} \vspace*{-1.0cm} \noindent }
\def\endwidebottom{
        \begin{multicols}{2} \vspace*{-1.0cm} \noindent }

\newcommand{\beq}{\begin{equation}}
\newcommand{\eeq}{\end{equation}}
\newcommand{\bdis}{\begin{displaymath}}
\newcommand{\edis}{\end{displaymath}}
\newcommand{\bea}{\begin{eqnarray}}
\newcommand{\eea}{\end{eqnarray}}
\newcommand{\barr}{\begin{array}}
\newcommand{\earr}{\end{array}}

\begin{document}

\title{Study of the connection between hysteresis and thermal relaxation in magnetic materials}

\author{Vittorio Basso, Cinzia Beatrice, Martino LoBue, Paola Tiberto and Giorgio Bertotti}

\address{Istituto Elettrotecnico Nazionale Galileo Ferraris and
        INFM, \\ Corso M. d'Azeglio 42, I-10125 Torino, Italy}

\maketitle

\begin{abstract}
The connection between hysteresis and thermal relaxation in
magnetic materials is studied from both the experimental and the
theoretical viewpoint. Hysteresis and viscosity effects are
measured in Finemet-type nanocrystalline materials above the Curie
temperature of the amorphous phase, where the system consists of
ferromagnetic nanograins imbedded in a paramagnetic matrix. The
hysteresis loop dependence on field rate, the magnetization time
decay at different constant fields and the magnetization curve
shape after field reversal are all consistent with a single value
of the fluctuation field $H_f \simeq $8Am$^{-1}$ (at
430$^{\circ}$C). In addition, it is shown that all data collapse
onto a single curve $M(H_{ath})$, when magnetization is plotted as
a function of a properly defined field $H_{ath}$, dependent on
time and field rate. Experimental data are interpreted by assuming
that the system consists of an assembly of elementary bistable
units, distributed in energy levels and energy barriers. The
approximations under which one predicts data collapse onto a
single curve $M(H_{ath})$ are discussed.
\end{abstract}

\date{\today}


\section{Introduction}

The joint presence of hysteresis and thermal relaxation is a
common situation in physical systems characterized by metastable
energy landscapes (magnetic hysteresis, plastic deformation,
superconducting hysteresis), and their interpretation still
represents a challenge to non-equilibrium thermodynamics.
Hysteresis is the consequence of the fact that, when the system is
not able to reach thermodynamic equilibrium during the time of the
experiment, the system will remain in a temporary local minimum of
its free energy, and its response to external actions will become
history dependent. On the other hand, the fact that the system is
not in equilibrium, makes it spontaneously approach equilibrium,
and this will give rise to relaxation effects even if no external
action is applied to the system.

In magnetic materials, thermal relaxation effects (also termed
magnetic after-effects or magnetic viscosity effects) are
particularly important in connection with data storage and with
the performance of permanent magnets, where a certain
magnetization state must be permanently conserved. Magnetic
viscosity experiments often show an intricate interplay with
hysteresis and with the role of field history in the preparation
of the system \cite{MULLER-96}. Thermal-activation-type models
have been proposed to interpret the fact that the initial stages
of relaxation often exhibit a logarithmic time decay of the
magnetization \cite{STREET-49}, but these models fail in
explaining the connection of viscosity to hysteresis properties.
Extensions were proposed to describe thermally activated dynamic
effects on hysteresis loops\cite{ESTRIN-89} and non-logarithmic
decay of the magnetization\cite{BROWN-96}. There exist in the
literature models for the prediction of coercivity, where thermal
activation over barriers plays a key role \cite{GIVORD-96}, but
these models usually do not pay attention to the problem of the
prediction of hysteresis under more complicated field histories.
An alternative is represented by detailed micromagnetic
descriptions of the magnetization process, coupled to Montecarlo
techniques for the study of its time evolution
\cite{KANAY-91,CHANTRELL-94}. In these cases, the extrapolation of
the results to long times and the identification of slow
(log-type) relaxation laws is far from straightforward.

Particular attention has been recently paid in the literature to
the joint description of hysteresis and thermal relaxation in
systems that are the superposition of elementary bistable units
\cite{BERTOTTI-96,MITCHELER-96}. The working hypothesis, inspired
by the results of several previous authors
\cite{NEEL-50,PREISACH-35,SOULETIE-83}, is that the free energy of
the system can be decomposed into the superposition of simple free
energy profiles, each characterized by two energy minima separated
by a barrier. The approach is able to predict, together with
hysteresis effects, the commonly observed logarithmic decay of the
magnetization at constant applied field as well as more
complicated history-dependent relaxation phenomena
\cite{LOBUE-97}. In this, it yields conclusions similar to those
given by hysteresis models driven by stochastic
input\cite{MAYERGOYZ-91}. When thermal activation effects are
negligible, the approach reduces to the Preisach model of
hysteresis, which provides quite a detailed description of several
key aspects of hysteresis \cite{MAYERGOYZ-BOOK}. The most
remarkable feature of the approach is that it yields this joint
description of hysteresis and thermal relaxation on the basis of a
few simple assumptions common to both aspects of the
phenomenology.

Following these considerations, in this article we investigate the
connection between hysteresis and thermal relaxation both from the
theoretical and the experimental viewpoint. Starting from the
general approach developed in Ref.\onlinecite{BERTOTTI-96} we
derived analytical laws for various types of history-dependent
relaxation patterns (Section II). The model predictions were then
applied to interpret experiments on a magnetic system particularly
suited to this task (Section III). The system, obtained by partial
crystallization of the amorphous
Fe$_{73.5}$Cu$_1$Nb$_3$Si$_{13.5}$B$_9$ alloy and known in the
literature as Finemet-type alloy, consists of a $\sim70\%$ volume
fraction of Fe-Si crystallites with diameters of about 10 nm,
imbedded in an amorphous matrix \cite{YOSHIZAWA-88}. The
crystalline and the amorphous phases are both ferromagnetic at
room temperature, where the system behaves like a good soft
magnetic material. However, the two phases have distinct Curie
temperatures, $T_c \simeq 350^{\circ}$C for the amorphous phase
and $T_c \simeq 700^{\circ}$C for the crystalline phase. When the
temperature is raised above $350^{\circ}$C and the amorphous phase
becomes paramagnetic, the grain-grain coupling provided by the
ferromagnetic matrix is switched off, and the system is
transformed into an assembly of magnetic nanograins randomly
dispersed in a non-magnetic matrix. This change results in a steep
increase of the coercive field, due to the decoupling of the
nanograins \cite{HERNANDO-94}, and in a definite enhancement of
thermal relaxation effects \cite{BASSO-98}, not only because the
temperature is increased, but also (and mainly) because the
typical activation volumes involved in magnetization reversal are
strongly reduced, again by the decoupling of the nanograins. This
is a situation where hysteresis and thermal relaxation acquire
comparable importance in determining the response of the system,
and where the time scale of relaxation effects becomes small
enough (in the range of seconds) to be amenable to a detailed
study.

We carried out a systematic experimental investigation of
hysteresis and thermal relaxation under various conditions, and we
made use of the theoretical approach of Section II to interpret
the experimental results. A remarkable general agreement between
theory and experiment was found. More precisely our main results
can be summarized as follows.

(i) The various relaxation patterns observed after different field
histories are all consistent with a unique value of the
fluctuation field \cite{CHANTRELL-94,WOHLFARTH-94}, $H_f =
k_BT/\mu_0M_sv$, where $k_B$ is the Boltzmann constant, $T$ is the
absolute temperature, $M_s$ is the saturation magnetization and
$v$ is the activation volume. At $T = 430 ^{\circ}$C, we found
$H_f \simeq$ 8 Am$^{-1}$. This corresponds to an activation volume
$v$ of linear dimensions $v^{1/3}$ of the order of 100 nm, which
indicates that, even well beyond the Curie point of the amorphous
matrix magnetization reversal still involves a consistent number
of coupled nanograins.

(ii) The saturation loop coercive field $H_c$ depends on the
applied field rate $dH/dt$ according to the law:

\beq
H_c = H_f  \ln(|dH/dt|) + C
\label{Hc1}
\eeq

\noindent where $H_f$ is the fluctuation field previously
mentioned and $C$ is a suitable constant.

(iii) Remarkable regularities are exhibited by the family of
relaxation curves $M(t;H_0,dH/dt)$, generated by starting from a
large positive field (positive saturation), then changing the
field down to the final value $H_0$ at the rate $dH/dt$, and
finally measuring the time decay of magnetization under the
constant field $H_0$. The family of experimental curves shows a
definite non-logarithmic behavior. However, all relaxation curves
collapse onto a single curve by plotting $M(t;H_0,dH/dt)$ as a
function of the athermal field $H_{ath}$, defined as

\beq H_{ath}(t;H_0,dH/dt) = H_0 \pm H_f \ln \left(
    \frac{t}{\tau_0} +
    \frac{H_f}{\tau_0 |dH/dt|}
\right)
\label{hath}
\eeq

\noindent where $\tau_0$ is a typical attempt time and the $\pm$
sign is the sign of the field rate $dH/dt$. This experimental
result represents an important confirmation of the theoretical
approach. In fact, the existence of the curve $M(H_{ath})$ is a
direct consequence of the fact that hysteresis and thermal
relaxation are controlled by the same distribution of energy
barriers. The curve $M(H_{ath})$ represents the magnetization
curve that one would measure if it was hypothetically possible to
switch off thermal effects completely and the athermal field
$H_{ath}$ plays the role of effective field summarizing the joint
effect of applied field and temperature.

(iv) When the external field is reversed at the turning point
$H_p$, the susceptibility $dM/dH$ after the turning point obeys
the law:

\beq
\frac{dM}{dH} = - \frac{\chi_{irr}}{2\exp(|H-H_p|/H_f)-1}
\label{susc}
\eeq

\noindent where $\chi_{irr}$ is the irreversible susceptibility
just before the turning point. This law is independent of the
field rate $dH/dt$.

(v) The role of field history is investigated by measurements of
the magnetization decay $M(t)$ under constant field $H_0$, carried
out after first decreasing the field from positive saturation down
to a certain reversal field $H_p < 0$, and then increasing it from
$H_p$ up to $H_0 > H_p$. Three distinct regimes emerge quite
distinctly: i) monotonic decrease of $M(t)$ for small $H_0-H_p$;
ii) monotonic increase for large $H_0-H_p$ ; iii) non monotonic
behavior in a intermediate region. We will show that also this
experimental behavior is in agreement with the predictions of the
model discussed in Section II.

The general conclusion of our analysis is that the individual
magnetization curves and relaxation laws can be quite complicated
and are not in general described by log-type laws. However, all of
them can be reduced to a small number of universal laws, which are
precisely the laws predicted by the model.

\section{Thermal activation in Preisach systems}

The presence of metastable states in the system free energy is the
key concept to understand hysteresis and thermal relaxation
effects. In this respect, the study of a system composed by a
collection of elementary bistable units \cite{BERTOTTI-BOOK}
yields substantial simplification without losing the basic
physical aspects of the problem. Each unit carries a magnetic
moment which can attain one of the two values $+\Delta m$ ("up" or
"+" state) and $-\Delta m$ ("down" or "-" state). The unit is
characterized by a simple double-well potential described by the
barrier height $\mu_0 h_c \Delta m$ and the energy difference
$2\mu_0 h_u \Delta m $ between the two states, where $h_c>0$ and
$h_u$ are field parameters characterizing the unit.

\begin{figure}[htb]
\narrowtext \centerline{
        \epsfxsize=4.0cm
        \epsfbox{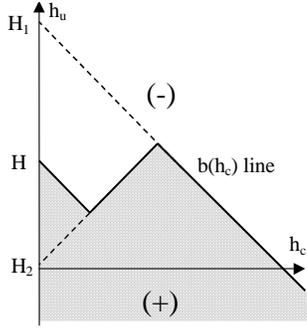}
        \vspace*{0.5cm}
        }
\caption{Preisach plane with example of state line $b(h_c)$
generated by the following field history: $-\infty$, $H_1$, $H_2$,
$H$} \label{fig:plane}
\end{figure}

Let us consider a system consisting of a collection of many such
units. The state of the collection is defined by specifying the
two subsets of units, S$_+$ and S$_-$, that are in the up and down
state at a certain time. At zero temperature the history of the
applied field only controls the shape of these subsets and the
description reduces to the Preisach model\cite{MAYERGOYZ-BOOK}. If
we represent each elementary unit as a point of the plane $(h_c,
h_u)$, known as Preisach plane, a certain field history will
produce a line $b(h_c)$ in the Preisach plane separating the S$_+$
and S$_-$ subsets (see Fig.\ref{fig:plane}).

The $b(h_c)$ line is just the set of internal state variables that
are needed to characterize the metastability of the system. In
this sense, the approach can be interpreted as a thermodynamic
formulation applicable to systems with hysteresis for which the
hypothesis of local equilibrium is not
valid\cite{BERTOTTI-96,BERTOTTI-BOOK}. All thermodynamic functions
can be expressed as functional of $b(h_c)$. In particular, the
magnetization $M$ is given by the integral \cite{MAYERGOYZ-BOOK}:

\beq
M = 2M_s \int_{0}^{\infty} dh_c \int_{0}^{b(h_c)}
p(h_c,h_u) dh_u
\label{PMintegral}
\eeq

\noindent where $M_s$ is the saturation magnetization and
$p(h_c,h_u)$ is the so-called Preisach distribution, giving the
statistical weight of each elementary contribution.
Eq.(\ref{PMintegral}) holds under the symmetry assumption
$p(h_c,-h_u) = p(h_c,h_u)$.

In presence of thermal activation, each unit $(h_c,h_u)$ relaxes
to its energy minimum, with the transition rates given by the
Arrhenius law. The interplay between external field changes and
thermal relaxation effects, once averaged over the entire
collection of units, determines the time evolution of the system.
When the system is far from equilibrium, the relaxation picture is
extremely complex and strongly history dependent. These aspects
have been extensively discussed in
Ref.\onlinecite{BERTOTTI-96,BERTOTTI-BOOK}. It has been shown
that, if the temperature is not too high, the relevant
contributions to the relaxation process are all concentrated
around the time-dependent state line $b(h_c,t)$, and the time
evolution of the state of the system is reduced to the time
dependence of the $b(h_c,t)$ line itself.  The following evolution
equation governs the state line:

\beq \frac{\partial b(h_c,t)}{\partial t}= 2
\frac{H_f}{\tau_0}\sinh \left[ \frac{H(t)-b(h_c,t)}{H_f} \right]
\exp \left[ - \frac{h_c}{H_f} \right]
\label{bhcequation}
\eeq

\noindent where $\tau_0$ is a typical attempt time, of the order
of 10$^{-9}$-10$^{-10}$ s, and $H_f=k_B T/\mu_0 \Delta m$ is the
so-called fluctuation field. In the limit $H_f \rightarrow 0$ the
effect of thermal activation vanishes and the solution of
Eq.(\ref{bhcequation}) yields the Preisach switching rules
\cite{BERTOTTI-BOOK}.

In order to make quantitative predictions about the magnetization
$M(t)$ (Equation(\ref{PMintegral})), one must know the system
state, given by the line $b(h_c,t)$, and the Preisach distribution
$p(h_c,h_u)$. The state line $b(h_c,t)$ can be derived, given the
field history $H(t)$, by solving Eq.(\ref{bhcequation}). We
consider here the case where the field history is composed of
arbitrary sequences of time intervals where the field stays
constant or the field varies at a given constant rate. In this
case Eq.(\ref{bhcequation}) can be exactly solved. Given a time
interval where $H(t)$ changes at a given constant rate $dH/dt$,
that is $H(t)=H_0+(dH/dt)t$, with initial conditions
$b(h_c,t=0)=b_0(h_c)$ and $H_0=b_0(0)$, one obtains:

\beq
b(h_c,t)=H(t)-2H_f \hbox{Arth}\Biggl[\mp 2
\frac{\tau_H}{\tau_c}+\frac{\tau_H}{\tau_s}\hbox{th}
\biggl[\pm\frac{t}{2\tau_s}+\hbox{Arth} \biggl(\pm 2
\frac{\tau_s}{\tau_c}+\frac{\tau_s}{\tau_H}\hbox{th}
\Bigl(\frac{H_0-b_0(h_c)}{2 H_f}\Bigr)\biggr)\biggr]\Biggr]
\label{solutiondhdt}
\eeq

\noindent where the upper (lower) sign correspond to positive
(negative) $dH/dt$, and

\bea
\tau_c=\tau_0\exp\left(\frac{h_c}{H_f}\right)
\label{tauc}\\
\tau_H=\frac{H_f}{|dH/dt|}
\label{tauh}\\
\tau_s=\frac{\tau_H\tau_c}{\sqrt{4\tau_H^2+\tau_c^2}}
\label{taus}
\eea

\noindent The special case where $H$ is constant in time, that is
$H(t)=H_0$, is obtained by taking the limit $dH/dt \rightarrow 0$,
in Eqs.(\ref{solutiondhdt})-(\ref{taus}).
Equation(\ref{solutiondhdt}) reduces to:

\beq
b(h_c,t)= H_0-2H_f\hbox{Arth}\left[
\hbox{th}\left(\frac{H_0-b_0(h_c)}{2H_f}\right)\exp\left(-\frac{2t}{\tau_c}\right)\right]
\label{solutionh0}
\eeq

\noindent The limit of Eq.(\ref{solutionh0}) for $t \rightarrow
\infty$ represents the equilibrium configuration at constant
field. One finds $b(h_c,\infty) = H_0$. On the other hand, the
limit of Eq.(\ref{solutiondhdt}) for $t \rightarrow \infty$ gives
the stationary state line under constant field rate, when all
transients related to the initial state $b_0(h_c)$ have died out.
One finds:

\beq
b(h_c,t)=H(t)\pm 2H_f \hbox{Arth} \left(2
\frac{\tau_H}{\tau_c}-\frac{\tau_H}{\tau_s}\right)
\label{solutionstat}
\eeq

For an arbitrary sequence of $n$ time intervals in which $H$
varies at different field rates, the resulting state line
$b(h_{c},t)$ is obtained by using the solution
(Eq.(\ref{solutiondhdt})) at the step $n-1$ as the initial
condition of step $n$.

\begin{figure}[htb]
\narrowtext \centerline{
        \epsfxsize=7.0cm
        \epsfbox{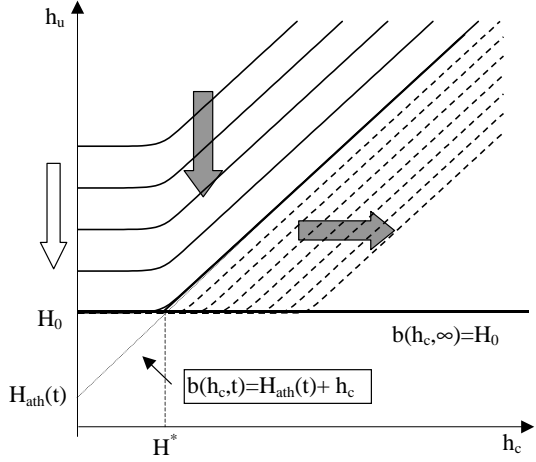}
        \vspace*{.5cm}
        }
\caption{State evolution for the field history considered in a
magnetic viscosity experiment. The system is prepared by starting
from positive saturation ($H \rightarrow \infty$), then bringing
the field down to the final value $H_0$ at the rate $dH/dt$, and
then, from the instant $t=0$, keeping $H_0$ constant over time.
Continuous lines: sequence of state lines $b(h_c,t)$ calculated
from Eq.(\ref{solutionstat}) in the case where the field $H(t)$
decreases from $+\infty$ at the field rate $dH/dt$. Dashed lines:
sequence of state lines $b(h_c,t)$ calculated from
Eq.(\ref{solutionh0}) under constant field $H_0$. As time
proceeds, $b(h_c,t)$ approaches the equilibrium state line
$b(h_c,\infty) = H_0$ by a front that moves at logarithmic speed.
In the region $h_c > H^*(t)$, the state line can be described as
$b(h_c,t) = H_{ath}(t)+h_c$, where the athermal field $H_{ath}(t)$
is given by Eq.(\ref{hath})} \label{fig:bhc1}
\end{figure}

Of particular interest is the field history commonly considered in
a magnetic viscosity experiment. The system is prepared by
starting from positive saturation ($H \rightarrow \infty$), then
bringing the field down to the final value $H_0$ at the rate
$dH/dt$, and then, from the instant $t=0$, keeping $H_0$ constant
over time. At $t=0$, the state line $b(h_c,0)$ is given by
Eq.(\ref{solutionstat}) (minus sign), with $H(0)=H_0$. The state
line describing the relaxation is obtained by inserting
Eq.(\ref{solutionstat}) as the initial condition of
Eq.(\ref{solutionh0}). The resulting $b(h_c,t)$ (see
Fig.\ref{fig:bhc1}) can be approximately divided into two parts:

\beq
b(h_c,t) = \left\{ \begin{array}{ll} H_0 & h_c<H^*(t)
\\ H_0-H^*(t)+h_c & h_c>H^*(t)
\end{array} \right.
\label{simple}
\eeq

\noindent where

\beq
H^*(t)=H_f\ln\left(\frac{\tau_H}{\tau_0}\right)+
H_f\ln\left(1+\frac{t}{\tau_H}\right)
\label{h*}
\eeq

\noindent At the initial time $t=0$ the state line is already
relaxed in the portion $h_c<H_f \ln(\tau_H/\tau_0)$ as a
consequence of the previous field history. Then the front
propagates at logarithmic speed and the final equilibrium state is
gradually approached.

As a conclusion to this section, we discuss a useful approximate
form of the results obtained so far. Let us consider the state
line of Eq.(\ref{simple}). The relaxed part, $b(h_c,t)=H_0$ for
$h_c<H^*(t)$, extends over an $h_c$ interval of a few times $H_f$
(see Eq.(\ref{h*})). Usually $H_f\ll H_c$, where $H_c$ represents
the coercive field. If the Preisach distribution is concentrated
around $h_c \sim H_c$, the contributions to the magnetization,
Eq.(\ref{PMintegral}), coming from the region $h_c<H^*(t)$ will be
small. Therefore, the magnetization calculated from
Eq.(\ref{PMintegral}) will not change substantially if one
modifies the true state line of Fig.\ref{fig:bhc1} into the line
$b(h_c,t)=H_{ath}(t)+h_c$ where $H_{ath}(t)=H_0-H^*(t)$. Perfectly
analogous considerations apply to the case where $H_0$ is reached
under positive $dH/dt$. The conclusion is that the magnetization
associated with different combinations of time and field rate will
be the same if one expresses the results in terms of the function
$M(H_{ath})$, where $H_{ath}$ is given by Eq.(\ref{hath}). The
field $H_{ath}$ plays the role of effective field summarizing the
effects of the applied field and of thermal activation. We stress
the fact that the existence of the function $M(H_{ath})$ is
independent of the details of the energy barrier distribution of
the system, provided the main approximation previously mentioned
is satisfied. The approximate law of corresponding states
expressed by $M(H_{ath})$ will be exploited in the analysis of the
experimental results presented in the next section.

\section{Thermal relaxation and hysteresis in nanocrystalline materials}

\subsection{Experimental setup}

We investigated the hysteresis properties of nanocrystalline
Fe$_{73.5}$Cu$_1$Nb$_3$Si$_{13.5}$B$_9$ (Finemet) alloys. This
material is commonly prepared by rapid solidification in the form
of ribbons approximately 20$\mu$m thick. The material is amorphous
in the as-cast state. Partial crystallization is induced by
subsequent annealing in furnace at 550$^\circ$C for 1h, with the
growth of Fe-Si crystal grains (with approximately 20 at \% Si
content). About 70\% of the volume fraction turns out to be
occupied by the Fe-Si crystal phase, in the form of nanograins of
about 10nm linear dimension, imbedded in the amorphous matrix. The
crystalline and the amorphous phases are both ferromagnetic, but
have quite distinct Curie temperatures: $T_c \simeq 350^{\circ}$C
for the amorphous phase, $T_c \simeq 700^{\circ}$C for the
crystalline phase. Therefore, above $350^{\circ}$C one has a
system composed of ferromagnetic nanograins imbedded in a
paramagnetic matrix, a situation in which the grain-grain coupling
is strongly reduced and relaxation phenomena become important.

The measurements were performed on a single strip (30cm long, 10mm
wide and 20$\mu$m thick) placed inside an induction furnace. The
temperature in the oven ranged from 20$^\circ$C to 500$^\circ$C,
always below the original annealing temperature. The sample, the
solenoid to generate the field and the compensated pick-up coils
were inserted in a tube kept under controlled Ar atmosphere. The
temperature, measured by a thermocouple, was checked to be
constant along the sample. The large thermal inertia of the
furnace permitted us to perform measurements under controlled
temperature with the heater off, in order to reduce electrical
disturbances. Experiments were performed up to a maximum
temperature of 500$^\circ$C and it was checked that no structural
changes were induced by the measurement at the highest
temperature. Experiments were performed under field rate $dH/dt$
in the range $10-10^6$ Am$^{-1}$s$^{-1}$. From 20$^\circ$C to
400$^\circ$C, we observed the increase of coercivity due to
magnetic hardening \cite{HERNANDO-98}. The paramagnetic transition
of the amorphous matrix causes a strong increase of the coercive
field and a decrease of the saturation magnetization. As expected,
after a peak around 400$^\circ$C, the coercivity decreases due to
the reduction of the Fe-Si anisotropy constant and the onset of
superparamagnetic effects. We selected, for our investigation, the
temperature $T=430^\circ$C, above the maximum coercivity, as the
point where nanograins are substantially decoupled. At this
temperature we measured thermal activation effects on:
\begin{itemize}
\item  saturation loop, that is: i) loops and coercivity versus field rate
and ii) relaxation curves versus applied field $H_0$ and field
rate $dH/dt$;
\item return branches with turning point $H_p$, that is: i) branch
shapes versus field rate and ii) relaxation curves versus field
history ($H_0$ and $H_p$).
\end{itemize}

\subsection{Thermal relaxation and dynamics along the saturation loop}

\subsubsection{$H_c$ vs. $dH/dt$}

\begin{figure}[htb]
\narrowtext \centerline{
        \epsfxsize=7.0cm
        \epsfbox{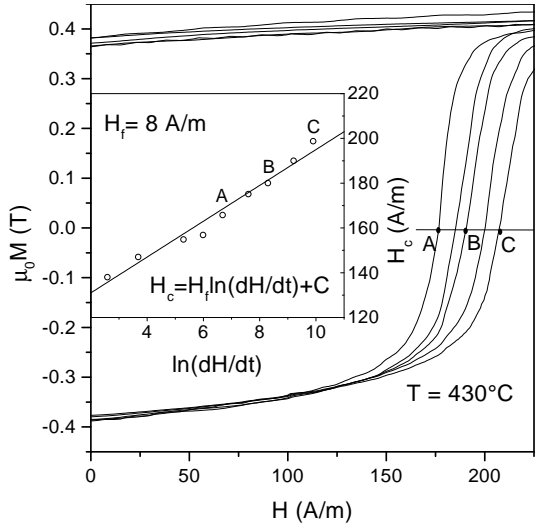}
        \vspace*{.5cm}
        }
\caption{Hysteresis loops measured on
Fe$_{73.5}$Cu$_1$Nb$_3$Si$_{13.5}$B$_9$ alloy (Finemet) at
T=430$^{\circ}$C under different field rates from 10
Am$^{-1}$s$^{-1}$ to 10$^{5}$ Am$^{-1}$s$^{-1}$. Inset:
Corresponding coercive field with best fit obtained from
Eq.(\ref{Hc1})} \label{fig:hc}
\end{figure}

We found that above $T \simeq 350^{\circ}$C hysteresis loop shapes
strongly depend on the field rate. Given the small ribbon
thickness and the high electrical resistivity of the alloy, this
dependence cannot be attributed to eddy current effects at least
for magnetizing frequencies below 100Hz. Fig.\ref{fig:hc} shows
hysteresis loops measured under different field rates at
$T=430^{\circ}$C. The inset shows the coercive field dependence on
field rate, together with the prediction of Eq.(\ref{Hc1}). Curve
fitting with $H_f$ as an adjustable parameter gives the result
$H_f$=8Am$^{-1}$.

Eq.(\ref{Hc1}) was found to be valid for the description of hard
magnetic materials \cite{GIVORD-96} and ultrathin ferromagnetic
films\cite{BRUNO-90}. In the model of Section II, the coercive
field $H_c$ is the field at which the state line $b(h_c)$ divides
the Preisach plane in two parts giving equal and opposite
contributions to the magnetization (Eq.(\ref{PMintegral})). When
the external field decreases from positive saturation at the
constant rate $dH/dt$, Eq.(\ref{solutionstat}) describes the
stationary regime where the state line $b(h_c,t)$ follows the
field at the same velocity and can be approximately divided into
two parts as in Eq.(\ref{simple}). When thermal activation is
negligible ($H_f \simeq 0$), the first part ($h_c<H^*$) is absent
and the coercive field $H_c \equiv H_c^i$ is the field at which
the line $b(h_c) = H_c^i + h_c$ gives $M=0$
(Eq.(\ref{PMintegral})). When thermal activation is important
($H_f \neq 0$) the state line is given by Eq.(\ref{simple}) and,
under the hypothesis that $p(h_c,h_u)$ is significantly different
from zero only in the region $h_c>H^*$ (see end of section II),
the zero magnetization state is given by the state line of
Fig.\ref{fig:bhc1} at $t=0$, $H_c=H_0$ and $H_{ath}(0)=H_c^i$.
Taking into account Eq.(\ref{hath}) (with $t=0$), we conclude that
the coercive field will depend on field rate according to the law:

\beq
H_c = H_c^i - H_f
\ln\left(\frac{\tau_H}{\tau_0}\right)
\label{Hc2}
\eeq

\noindent where $\tau_H$ is given by Eq.(\ref{tauh}). By assuming
$\tau_0 \sim 10^{-10}$s, we obtain from the data of
Fig.\ref{fig:hc} $H_c^i \simeq 320$Am$^{-1}$. At the lowest
measured field rate $dH/dt$= 13.3 Am$^{-1}$s$^{-1}$, we have, from
Eq.(\ref{h*}), that the state line is relaxed up to $h_c \simeq
180$Am$^{-1}$. By using Eq.(\ref{Hc2}), one can derive the limit
field rate at which thermal effects become unimportant as $dH/dt =
H_f/\tau_0$ = 8 10$^{10}$ Am$^{-1}$s$^{-1}$; and the
superparamagnetic limit, where the coercive field vanishes, as
$dH/dt$ = 5.6 10$^{-7}$ Am$^{-1}$s$^{-1}$.

\subsubsection{Relaxation vs. $H_0$ and $dH/dt$}

\begin{figure}[htb]
\narrowtext \centerline{
        \epsfxsize=8.0cm
        \epsfbox{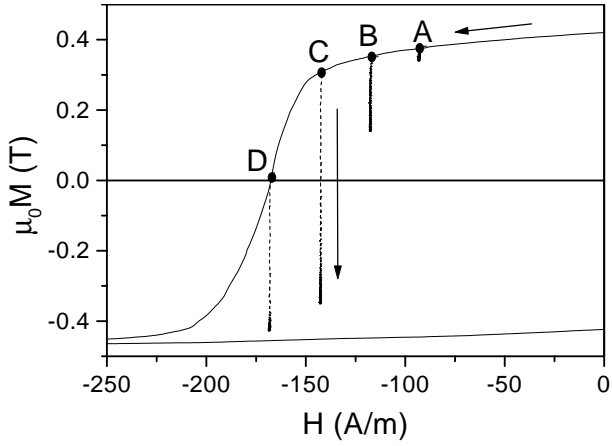}
        \vspace*{.5cm}
        }
\caption{Magnetization decay under constant field $H_0$ measured
on Finemet at T=430$^{\circ}$C. The different fields $H_0$ (-93
Am$^{-1}$ (A); -118 Am$^{-1}$ (B); -143 Am$^{-1}$ (C); -168
Am$^{-1}$ (D)) are reached under identical field rate $dH/dt=6.25
10^{3}$Am$^{-1}$s$^{-1}$. The figure shows the extent of decay in
0.5 s (see Fig.\ref{fig:relt}).} \label{fig:relHM}
\end{figure}

\begin{figure}[htb]
\narrowtext \centerline{
        \epsfxsize=8.0cm
        \epsfbox{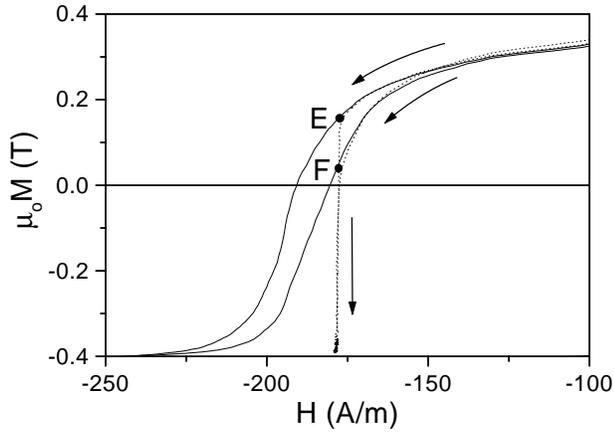}
        \vspace*{.5cm}
        }
\caption{Magnetization decay under constant field $H_0$ measured
on Finemet at T=430$^{\circ}$C. The same field $H_0=-178$Am$^{-1}$
is reached at different field rates $dH/dt$ (2
10$^4$Am$^{-1}$s$^{-1}$ (E); 10$^{4}$Am$^{-1}$s$^{-1}$ (F)). The
figure shows the extent of decay in 0.01 s. (see
Fig.\ref{fig:relt})} \label{fig:reldHM}
\end{figure}

\begin{figure}[htb]
\narrowtext \centerline{
        \epsfxsize=8.0cm
        \epsfbox{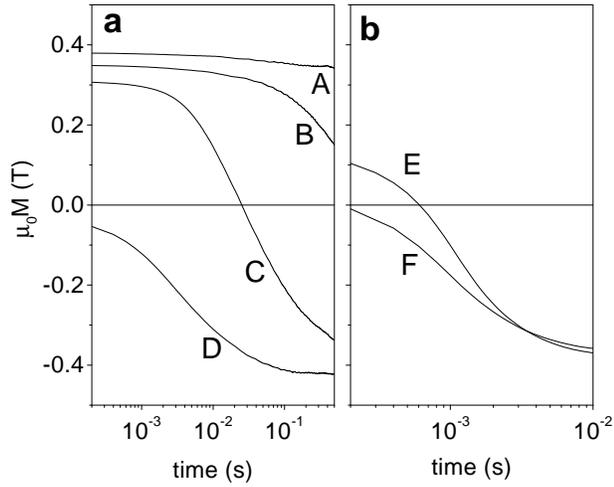}
        \vspace*{.5cm}
        }
\caption {Time behavior of the magnetization $M(t)$ for the field
histories of Fig.\ref{fig:relHM} (left) and Fig.\ref{fig:reldHM}
(right).} \label{fig:relt}
\end{figure}

\begin{figure}[htb]
\narrowtext \centerline{
        \epsfxsize=8.0cm
        \epsfbox{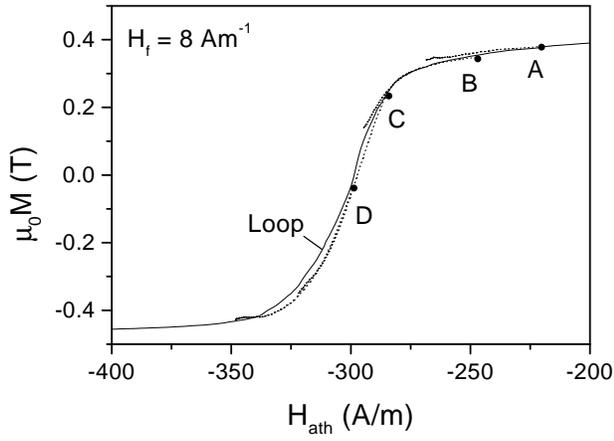}
        \vspace*{.5cm}
        }
\caption{Relaxation curves of Fig.\ref{fig:relt}a (broken lines)
 and hysteresis loop of Fig.\ref{fig:relHM} (solid line) as a
function of the field $H_{ath}$ of Eq.(\ref{hath}), calculated
assuming $H_f=8$Am$^{-1}$ and $\tau_0=10^{-10}$ s}
\label{fig:relHath}
\end{figure}

The relaxation experiment is performed by applying a large
positive field, which is then decreased at a fixed rate $dH/dt$ to
the final negative value $H_0$. The magnetization is then measured
as a function of time, under constant $H_0$. We performed a
systematic study of the relaxation behavior by changing $H_0$ and
$dH/dt$. In general, we found that thermal relaxation results in
large non-logarithmic variations of the magnetization.
Figs.\ref{fig:relHM},\ref{fig:reldHM},\ref{fig:relt} show i) the
relaxation at different fields reached under the same field rate
and ii) the relaxation at the same field, when $H_0$ is reached at
different field rates. All relaxation curves collapse onto a
single curve by plotting $M$ versus $H_{ath}$, given by
Eq.(\ref{hath}) (see Fig.\ref{fig:relHath}). To obtain the
$M(H_{ath})$ curve, the only parameter to be set is the
fluctuation field $H_f$. Data collapse onto a unique curve by
assuming $H_f=8$Am$^{-1}$. The same curve collapse was found to be
valid for the loops of Fig.\ref{fig:hc}, when plotted as a
function of the athermal field $H_{ath}$, with $t=0$. As an
example, Fig.\ref{fig:relHath} shows the result obtained for the
loop measured at $dH/dt$=6.25 10$^3$ Am$^{-1}$s$^{-1}$
(Fig.\ref{fig:relHM}), again assuming $H_f=8$Am$^{-1}$.

These regularities can be derived under the Preisach description
of the system by the approximations discussed at the end of
Section II, that is, by assuming that the Preisach distribution is
significantly different form zero only in the region $h_c>H^*$.
The field $H_{ath}$ plays the role of effective driving field,
summarizing the effect of applied field and thermal activation.
This conclusion supports the idea that hysteresis and thermal
activation phenomena depend on the same distribution of energy
barriers.

\subsection{Thermal relaxation and dynamics along return
branches}

\subsubsection{Return branches vs. $dH/dt$}
\begin{figure}[htb]
\narrowtext \centerline{
        \epsfxsize=8.0cm
        \epsfbox{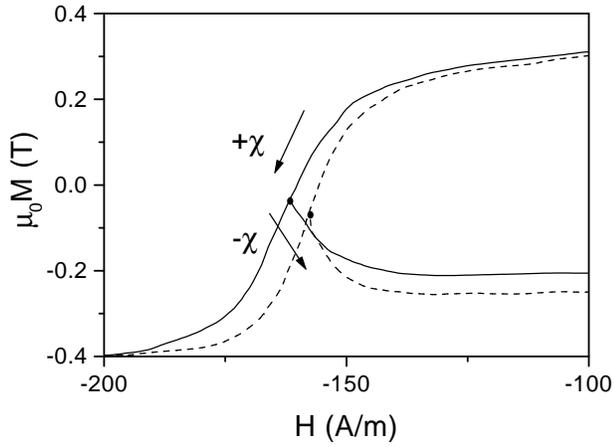}
        \vspace*{0.5cm}
        }
\caption{Return branches measured on Finemet at T=430$^{\circ}$C
after field reversal at $H=H_p$. Continuous lines: $dH/dt$=2
10$^3$ Am$^{-1}$s$^{-1}$, $H_p=-160.5$ Am$^{-1}$; broken lines:
$dH/dt$=8 10$^2$Am$^{-1}$s$^{-1}$, $H_p=-157.5$ Am$^{-1}$.}
\label{fig:chihist}
\end{figure}

The role of field history on hysteresis curves was investigated by
the measurement of recoil branches. We observed that when the
field is reversed at the turning point $H=H_p$, the differential
susceptibility after the reversal point is initially negative and
equal to the susceptibility just before the turning point (see
Fig.\ref{fig:chihist}). This effect is found to be independent of
the field rate. After the turning point, the negative
susceptibility decays to zero in a field interval of the order of
the fluctuation field $H_f$. This effect was observed for several
temperatures and peak field amplitudes.

\begin{figure}[htb]
\narrowtext \centerline{
        \epsfxsize=8.0cm
        \epsfbox{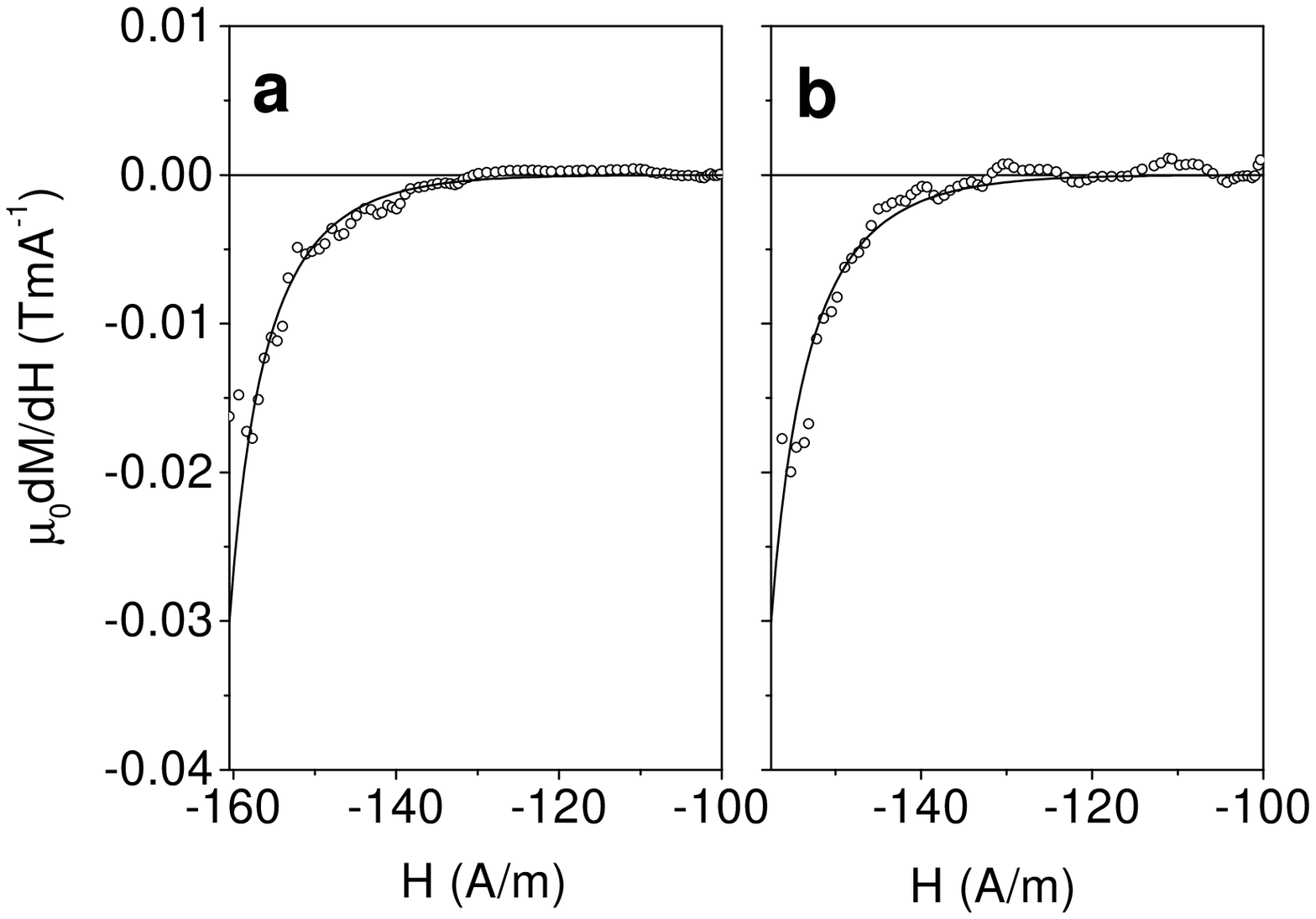}
        \vspace*{0.5cm}
        }
\caption{Differential susceptibilities measured on Finemet at
T=430$^{\circ}$C after a turning point at $H=H_p$. Left: $dH/dt=2
10^3$Am$^{-1}$s$^{-1}$, $H_p=-160.5$ Am$^{-1}$. Right: $dH/dt=8
10^2$Am$^{-1}$s$^{-1}$, $H_p=-157.5$ Am$^{-1}$. (see
Fig.\ref{fig:chihist}). Continuous lines: Eq.(\ref{susc}) with
$H_f=8$Am$^{-1}$.} \label{fig:chi}
\end{figure}

In order to explain this behavior, let us consider the system
state in the Preisach plane. The $b_T(h_c)$ line correspondent to
the turning point is given by Eq.(\ref{solutionstat}) with
$H_0=H_p$. When the field $H$ is increased after the turning
point, $b(h_c)$ is given by Eq.(\ref{solutiondhdt}), with
$dH/dt>0$ and $b_T(h_c)$ as the initial condition. The resulting
solution for $b(h_c,t)$(see Fig.\ref{fig:bhc2}) shows that after
the turning point a part of the state line still moves downward
even if the field is increasing. This part of the line can be
approximately described as $b(h_c) = H^+ + h_c$,where:

\beq
H^+ = H_p - H_f \left[ \ln\left( \frac{\tau_H}{\tau_0}\right)
+ \ln\left(2-\exp\left(- \frac{|H-H_p|}{H_f}\right) \right)\right]
\label{h+}
\eeq

\noindent where the $\pm$ is the sign of $dH/dt$ before the
turning point. Under the approximation described at the end of
Section II, i.e. that the Preisach distribution is concentrated at
$h_c>H^*$, the susceptibility after the turning point is obtained
by inserting $b(h_c)=H^+ + h_c$ into Eq.(\ref{PMintegral}) and
taking the first derivative with respect to $H$. The result is
given by Eq.(\ref{susc}), where $\chi_{irr} =2 M_s \int
p(h_c,b(h_c))dh_c$ is the irreversible susceptibility before the
turning point and the dependence on $|dH/dt|$ disappears. The fit
of Eq.(\ref{susc}) to experimental data is shown in
Fig.\ref{fig:chi}, where the only free parameter $H_f$ is found to
be $\simeq$8 A/m, coherently with the other results previously
discussed.

\subsubsection{Relaxation curves vs. field history ($H_0$ and $H_p$)}
The role of field history on the relaxation effects was
investigated by measuring the time decay of $M(t)$ at the field
$H_0$ applied after the turning point $H_p$. In the case $H_p < 0$
and $H_0 > H_p$ we found tree distinct regimes: i) monotone
decrease of $M(t)$ for small $H_0-H_p$; ii) monotone increase for
large $H_0-H_p$ ; iii) non monotone behavior in a intermediate
region.

\begin{figure}[htb]
\narrowtext \centerline{
        \epsfxsize=6.0cm
        \epsfbox{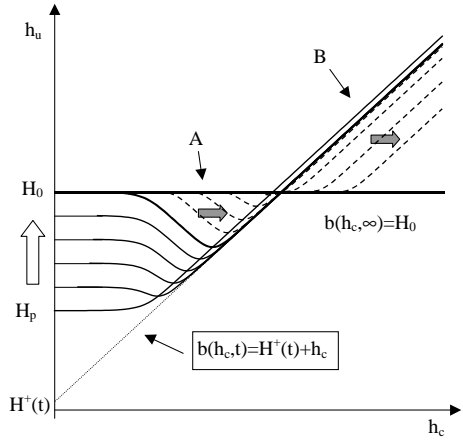}
        \vspace*{0.5cm}
        }
\caption{State evolution after field reversal at $H=H_p$.
Continuous lines: when $H$ increases from $H_p$ to $H_0$ the state
line $b(h_c,t)$ is given by Eq.(\ref{solutiondhdt}), with
Eq.(\ref{solutionstat}) as initial condition. Part B of the state
line moves downward even if the field is increasing. Dashed lines:
once the field has reached the constant value $H_0$ the time
dependence of the state line is given by Eq.(\ref{solutionh0}).
Relaxation proceeds through the motion of the two fronts indicated
by the gray arrows. Part B of the state line can be described as
$b(h_c,t) = H^+(t)+h_c$, where the $H^+(t)$ is given by
Eq.(\ref{H+t}).} \label{fig:bhc2}
\end{figure}

These three regimes are predicted by the model of Section II.
Fig.\ref{fig:bhc2} shows that part A and B relax at logarithmic
speed toward equilibrium with two different time constants and
give contributions to the magnetization of opposite sign. However,
quantitative predictions need a detailed knowledge of the Preisach
distribution shape. We limit here our analysis to the case i)
where, the contribution of the front A of Fig.\ref{fig:bhc2} is
small. In the region $h_c>H^*$ one finds that the state line can
be approximately described as $b(h_c)=H^+(t)+h_c$, where

\beq H^+(t)=H^+(0)-H_f \ln\left(
1+\frac{t}{\tau_H\left(2\exp\left(|H_0-H_p|/H_f\right)-1\right)}\right)
\label{H+t}
\eeq

\noindent and $H^+(0)$ is given by Eq.(\ref{h+}). Since the system
state can be identified by $H^+(t)$, this field assumes the same
role of the athermal field of Section III.B.ii).  By plotting a
relaxation curve $M(t)$ measured at $H_0=-153$Am$^{-1}$,
$H_p=-163$Am$^{-1}$ and $dH/dt=2 10^3$Am$^{-1}$s$^{-1}$ as a
function of $H^+(t)$ of Eq.(\ref{H+t}) we found that, with
$H_f=8$Am$^{-1}$, the resulting curve collapse on the $M(H_{ath})$
of Fig.\ref{fig:relHath}.

\section{Conclusions}

We have studied hysteresis and magnetic relaxation effects in
Finemet-type nanocrystalline materials above the Curie temperature
of the amorphous matrix, where the system consists of
ferromagnetic nanograins ($\simeq $10 nm linear size) imbedded in
a paramagnetic matrix. This is a situation where hysteresis and
thermal relaxation acquire comparable importance in determining
the response of the system, and where the time scale of relaxation
effects becomes small enough (in the range of seconds) to be
amenable to a detailed study. Experiments have been carried out by
investigation of the hysteresis loops dependence on the field
rate, the magnetization time decay at different constant fields
and the magnetization curve shape after field reversal. It is
shown that all the experimental data can be explained by a model
based on the assumption that the system consists of an assembly of
elementary bistable units, distributed in energy levels and energy
barriers. This approach permits one to describe all the measured
effects in terms of a single parameter, the fluctuation field
$H_f$, that was found to be $H_f \simeq $8Am$^{-1}$ (at
430$^{\circ}$C). This corresponds to an activation volume $v$ of
linear dimensions $v^{1/3}$ of the order of 100 nm, which
indicates that, even well beyond the Curie point of the amorphous
matrix, magnetization reversal still involves a consistent number
of coupled nanograins \cite{HERNANDO-94}. In addition, the joint
effect of the applied field and the thermal activation can be
summarized by an effective field $H_{ath}$, and the measured
curves can be rescaled onto to a single curve $M(H_{ath})$. The
existence of the $M(H_{ath})$ curve, which is a direct prediction
of the model, strongly support the idea that hysteresis and
thermal relaxation are controlled by the same distribution of
energy barriers. The results here obtained may represent a general
framework for the study of the connection between hysteresis and
thermal relaxation in different systems, such as materials for
recording media and permanent magnets.



\newpage

\end{document}